\documentclass[aps,prl,prabib,epsfig,amsmath,amssymb,twocolumn,superscriptaddress,showpacs]{revtex4}
\usepackage{amsmath}
\usepackage{amssymb}
\usepackage{epsfig}
\usepackage{graphics}

\newcommand \beq{\begin{eqnarray}}
\newcommand \eeq{\end{eqnarray}}
\def\simge{\mathrel{%
       \rlap{\raise 0.511ex \hbox{$>$}}{\lower 0.511ex \hbox{$\sim$}}}}
\def\simle{\mathrel{
       \rlap{\raise 0.511ex \hbox{$<$}}{\lower 0.511ex \hbox{$\sim$}}}}

\newcommand{\rr}{\mathbf{r}}
\newcommand{\dd}{\mathbf{d}}

\begin{document}

\title{Clock shifts of optical transitions in ultracold atomic gases}

\author{Zhenhua Yu}
\affiliation{The Niels Bohr International Academy, The Niels Bohr Institute, University of Copenhagen, Blegdamsvej 17, DK-2100 Copenhagen \O, Denmark}
\author{C.\ J.\  Pethick}
\affiliation{The Niels Bohr International Academy, The Niels Bohr Institute, University of Copenhagen, Blegdamsvej 17, DK-2100 Copenhagen \O, Denmark}
\affiliation{ NORDITA, Roslagstullsbacken 21, 10691 Stockholm, Sweden}


\begin{abstract}
We calculate the shift, due to interatomic interactions, of an
optical transition in an atomic Fermi gas trapped in an optical
lattice, as in recent experiments of Campbell {\it et al.}, Science
{\bf 324}, 360 (2009).  Using a pseudospin formalism to describe the
density matrix of the internal two states of the optical transition,
we derive a Bloch equation which incorporates both the spatial
inhomogeneity of the probe laser field and the interatomic
interactions. Expressions are given for the frequency shift as a
function of the pulse duration, detuning of the probe laser, and the
spatial dependence of the electric field of the probe beam.  In the
low temperature semiclassical regime, we find that the magnitude of
the shift is proportional to the temperature.
\end{abstract}

\maketitle In the continuing quest to develop improved atomic
clocks, attention has turned to fermionic atoms because shifts of transition frequencies
due to interatomic interactions (so-called clock shifts)
are expected to be strongly suppressed in gases of identical fermions due to the Pauli exclusion
principle. It came as a surprise that nonzero interaction shifts were observed in a recent experiment on the
$^1$S$_0$--$^3$P$_0$ optical transition for $^{87}$Sr atoms trapped in an optical lattice  \cite{campbell, ye}, since for a gas of fermions it has been shown theoretically that for
a homogeneous probe field
there should be no frequency
shift \cite{su2}. Thus
inhomogeneity is indispensable for observing nonzero frequency shifts
in identical fermion samples.
The authors of Ref.\ \cite{campbell}
attributed the shifts to the combined influence of the spatial
inhomogeneity of the probe laser field and the interatomic
interaction.  The same underlying physical mechanism gives
rise to the Leggett-Rice effect in spin transport in liquid $^3$He
\cite{leggett}, and spin segregation in ultracold bosons
\cite{cornell} and fermions \cite{thomas}.

In this paper, we introduce pseudospin operators to describe atomic
correlations within each motional states of the atoms and derive a
Bloch equation that describes the evolution of the pseudospin under
the combined effect of an external probe laser field and interatomic
interactions and derive expressions for the frequency shifts to be
expected under the experimental conditions of  Ref.~\cite{campbell}.
Since this work was largely completed, we became aware of
Refs.\  \cite{gibble, rey} in which, for a particular form of the
inhomogeneity of the probe field, the problem is approached using
the wave function of the state.

{\it Basic formalism.} In the experiment, atoms of $^{87}$Sr, a fermionic isotope, were
initially prepared in one hyperfine state of the $^3$P$_0$
excited-state manifold (denoted by $e$) and transferred to a
hyperfine state of the $^1$S$_0$ ground-state manifold (denoted by
$g$) by application of a probe laser field \cite{campbell}. In the
absence of the laser field, the system is described by the Hamiltonian
\begin{align}
H=& \int d^3r \sum_{\alpha=e,\,g}
 \psi_\alpha^\dagger(\rr)\left[-\frac{\nabla^2}{2m}+E_\alpha+V(\rr)\right]
 \psi_\alpha(\rr)\nonumber\\
 &+g\int d^3r
 \psi_e^\dagger(\rr)\psi_g^\dagger(\rr) \psi_g(\rr) \psi_e(\rr),
 \label{ham}
\end{align}
where the $\psi_\alpha$ are annihilation operators for atoms in the
internal states $|\alpha\rangle$ and $E_\alpha$ are the internal
energies. The effective low energy interaction coupling is given by
$g=4\pi a_s/m$ ($\hbar=1$) where $a_s$ is the s-wave scattering
length for collisions between a ground state atom and an excited
state one. The external dipole optical potential $V(\rr)$ is a
superposition of a deep one-dimensional optical lattice potential
aligned along the $z$ direction for which the frequency of
oscillations about the minima of the potential is $\omega_{\rm lattice} \approx 80$
kHz, and a harmonic potential with
$\omega_{\perp}=\omega_{x}=\omega_y\approx450$ Hz and $\omega_z\ll 80$
kHz. (The potential $V(\rr)$ is the same for the two atomic states since the wavelength of the optical lattice is chosen to be
$\lambda_{\rm lattice} \approx 813.429$ nm, the magic wavelength at which the
electric polarizabilities are the same for the ground and the
excited states.)  Atoms are thereby confined to a number of pancake-shaped regions extended in the $x$- and $y$-directions and centered on the $z$-axis, and are essentially limited to the ground state for motion in the $z$-direction under experimental conditions.

 It is convenient to introduce the pseudospin
density operators $\hat{\bf
S}(\rr)=[\psi^\dagger_e(\rr),\psi^\dagger_g(\rr)]{\boldsymbol
\sigma} [\psi_e(\rr),\psi_g(\rr)]^T/2$, where $\boldsymbol  \sigma$
are the Pauli matrices. The coupling between the probe laser field
$\mathbf E(\rr,t)$ and the atoms is due to dipole transitions
between the two states, and can be described by the Hamiltonian
\begin{align}
H_{\rm probe}&=-\int d^3r \left(\langle e|\dd \cdot{\bf E}(\rr, t)|g\rangle\,\psi_e^\dagger(\rr) \psi_g(\rr)
+ {\rm h.c.}\right) \nonumber \\
&=-\int d^3r\left({\cal B}_x(\rr, t) {\hat S}_x(\rr) +{\cal
B}_y(\rr, t) {\hat S}_y(\rr)\right)
\end{align}
where ${\cal B}_x(\rr, t)=\langle e|\mathbf d\cdot\mathbf
E(\rr,t)|g\rangle+\langle g|\mathbf d\cdot\mathbf E(\rr,t)|e\rangle$
and ${\cal B}_y(\rr, t)=i(\langle e|\mathbf d\cdot\mathbf
E(\rr,t)|g\rangle-\langle g|\mathbf d\cdot\mathbf
E(\rr,t)|e\rangle)$ are the components of the ``pseudomagnetic
field" \cite{Directions}, and $\mathbf d$ is the electric dipole
moment. Note that Refs.~\cite{gibble,rey} assume only $\mathcal
B_x(\rr,t)$ nonzero. However, both $\mathcal B_x(\rr,t)$ and
$\mathcal B_y(\rr,t)$ depend on the details of the probe laser field
and are generally not zero.

{\it Bloch equation.}
We study the coherent evolution of the system governed by
$H+H_{\rm probe}$ through the Bloch equations for the pseudospin
operators. We expand the field operators $\psi_\alpha(\rr)=\sum_i a_{\alpha
i}\phi_i(\rr)$, where $\phi_i(\rr)$ are the free particle
eigenstates for $H$ with $g=0$, and $a_{\alpha i}$ are the
corresponding annihilation operators. In the frame co-rotating with
the probe laser, we derive
\begin{align}
\frac{d}{dt}\mathbf{S}_i=\mathbf{\Omega}_i\times\mathbf{S}_i+2\sum_j
g_{ij}\mathbf{S}_i\times\mathbf{S}_j,\label{bloch}
\end{align}
where $\mathbf S_i=\langle [a^\dagger_{ei},
a^\dagger_{gi}]\boldsymbol\sigma[a_{ei}, a_{gi}]^T\rangle/2$ and
$g_{ij}=g\int d^3r|\phi_i(\mathbf{r})\phi_j(\mathbf{r})|^2$. Thus
the expectation value of the total pseudospin is given by $\mathbf
S=\sum_i\mathbf S_i$  with $\langle\dots\rangle$ denoting the
trace with the density matrix. The probe laser field has a
generic form $\mathbf E(\mathbf r,t)=\mathbf E_-(\mathbf
r)\exp(-i\omega_L t)+\mathbf E^*_-(\mathbf r)\exp(i\omega_L t)$ with
$\omega_L$ the laser frequency. The driving fields
$\mathbf{\Omega}_i=(\Omega_{ix},\Omega_{iy},\Omega_{iz})$ have
components $\Omega_{ix}=\Omega_{i-}+{\rm c.c.}$,
$\Omega_{iy}=i(\Omega_{i-}-{\rm c.c.})$ and $\Omega_{iz}=\Delta$,
where $\Omega_{i-}=-\int d^3r \, |\phi_i(\rr)|^2\langle
 e|\dd\cdot\mathbf E_-(\rr)|g\rangle$ and $\Delta=E_e-E_g-\omega_L$  is the
detuning.
Equation (\ref{bloch}) is valid in the weak interaction limit, $g\to
0$, and the Lamb-Dicke regime where the recoil of the atoms due to
the scattering with the probe laser is negligible and it
shows that the magnitude of the pseudospin $S_i$ remains unaltered.

At this stage,
the joint effect of the spatial inhomogeneity of $\mathbf E(\rr,t)$
and the interaction is manifest in Eq.~(\ref{bloch}): if all
pseudospins $\mathbf S_i$ initially point towards the north pole of the
Bloch sphere (as in Ref.~\cite{campbell}), corresponding to all atoms being in the excited state, the first term causes the pseudospins of states with
different $\mathbf\Omega_i$ to precess by different amounts, and
then the interatomic interaction term no longer vanishes.

We convert Eq.~(\ref{bloch}) into an
integral equation,
\begin{align}
\mathbf{S}_i(t)=&\mathbf{G}_i(t)\mathbf{S}_i(0)\nonumber\\
&+2\sum_{j} g_{ij}\int^t_0 dt'\;
\mathbf{G}_i(t-t')\mathbf{S}_i(t')\times\mathbf{S}_j(t') ,
\label{integral}
\end{align}
where $\mathbf{S}_i(0)$ are the initial values of the pseudospins. In the
Cartesian basis ($x,y,z$), the Green's function satisfies the equation
\begin{align}
\left(\frac{d}{dt}-\mathbf Q\right)\mathbf
G(t)=\delta(t)\label{greene}
\end{align}
with
\begin{eqnarray}
\mathbf Q=\Omega\left[
 \begin{array}{ccc}
     0 & -\cos\theta & \sin\theta\sin\phi\\
\cos\theta & 0 & -\sin\theta\cos\phi\\
-\sin\theta\sin\phi & \sin\theta\cos\phi & 0
 \end{array}\right],
\end{eqnarray}
where $\Omega_x=\Omega\sin\theta\cos\phi$,
$\Omega_y=\Omega\sin\theta\sin\phi$, $\Omega_z=\Omega\cos\theta$ and
$\Omega=\sqrt{\Omega_x^2+\Omega_y^2+\Omega_z^2}$. The matrix
$\mathbf Q$ can be diagonalized by the matrix
$\mathbf A=(v_1,v_2,v_3)$ where the vectors are given by
$v_1^{T}=(\sin\theta\cos\phi,\sin\theta\sin\phi,\cos\theta)$,
$v_2^T=(\cos\theta\cos\phi+i\sin\phi,\cos\theta\sin\phi-i\cos\phi,-\sin\theta)/\sqrt2$
and
$v_3^T=(\cos\theta\cos\phi-i\sin\phi,\cos\theta\sin\phi+i\cos\phi,-\sin\theta)/\sqrt2$.
Thus
\begin{align}
\mathbf G(t)=\theta(t)\mathbf A\begin{bmatrix}1&0&0\\
0& e^{i\Omega t} &0\\
0&0& e^{-i\Omega t}\end{bmatrix}\mathbf A^\dagger. \label{greens}
\end{align}

From Eq.~(\ref{integral}), the deviation of the total pseudospin $\mathbf
S$ from its value without interaction, $\mathbf S^0(t)=\sum_i\mathbf
G_i(t)\mathbf S^0_i(0)$, is given by
\begin{align}
&\delta\mathbf{S}(t)\nonumber\\
&=\sum_{i,j}g_{ij}\int^t_0
dt' \left[\mathbf{G}_i(t-t')-\mathbf G_j(t-t')\right]\left[
\mathbf{S}_i(t')\times\mathbf{S}_j(t')\right], \label{ds}
\end{align}
where the subscript
$i$ of $\mathbf G$ indicates $\mathbf\Omega=\mathbf\Omega_i$.
Equation (\ref{ds}) shows that the change in the pseudospin is zero if $\mathbf\Omega_i$ is independent of $i$.

{\it Frequency shift.} In the experiments the frequencies
$\delta\mathbf\Omega_{ij}=(\mathbf \Omega_i- \mathbf\Omega_j)/2$ and
$g_{ij}$ are small compared with the average $\mathbf \Omega_i$ and
therefore we may treat them perturbatively.  For the interatomic
interaction this means that the change in the total pseudospin is a
sum over contributions from independent pairs of motional states,
and therefore we may consider just two motional states (which we
shall denote by 1 and 2), and then sum over all possible pairs at
the end. From Eq.~(\ref{bloch}), the total pseudospin of the pair of
states $\mathbf S_+=\mathbf S_1+\mathbf S_2$ and the difference
$\mathbf S_-=\mathbf S_1-\mathbf S_2$ satisfy the equations
\begin{align}
\frac{d}{dt}\mathbf{S}_+=&\mathbf{\Omega}\times\mathbf{S}_++\delta\mathbf\Omega\times\mathbf
S_-
\label{twoplus}\\
\frac{d}{dt}\mathbf{S}_-=&\mathbf{\Omega}\times\mathbf{S}_-+\delta\mathbf\Omega\times\mathbf
S_+ -2g_{12}\mathbf{S}_+\times\mathbf{S}_-, \label{twominus}
\end{align}
with $\mathbf \Omega=(\mathbf\Omega_1+\mathbf\Omega_2)/2$ and
$\delta\mathbf\Omega=(\mathbf\Omega_1-\mathbf\Omega_2)/2$. Note the
interaction does not appear in the equation for $\mathbf S_+$
explicitly. This is consistent with the argument based on the SU(2)
symmetry of the interaction \cite{su2}: if $\delta\mathbf\Omega=0$,
the dynamics of $\mathbf S_+$ is only governed by $\mathbf\Omega$
and no interaction effects can show up. The physical mechanism for
creating a frequency shift due to interatomic interactions is as
follows.  Initially all pseudospins are aligned in the same
direction, and therefore $\mathbf{S}_+\times\mathbf{S}_-$ is zero
and the interaction has no effect. However, according to Eq.\
(\ref{twominus}) the field inhomogeneity creates a component of
$\mathbf{S}_-$  perpendicular to $\mathbf{S}_+$, and therefore the
interactions can have an effect.  Subsequently, according to Eq.\
(\ref{twoplus}) this results in a change in  $\mathbf{S}_+$.
Consequently, the leading contribution to the shift is proportional
to $g (\delta \Omega)^2 $.

For the initial conditions
of interest, $\mathbf S_1(0)=S_1(0)\mathbf e_z$  and $\mathbf
S_2(0)=S_2(0)\mathbf e_z$, where $S_i(0)=f_i/2$ and $f_i$ is the
initial distribution function.  Solving for the change of $\mathbf S_+(t)$
due to the interaction to second order in $\delta\mathbf \Omega$ and
first order in $g_{12}$ by iterating integral equations similar to
Eq.~(\ref{integral}), we obtain
\begin{align}
&\delta \mathbf S_+(t)\nonumber\\
=&2\int_0^t dt' \int_0^{t'}
dt'' \int_0^{t''}dt'''\mathbf
G(t-t')\delta\mathbf\Omega\times\mathbf G(t'-t'')
\nonumber\\
&g_{12}\Big\{\big[\mathbf G(t'')\mathbf S_-(0)\big]\times\big[\mathbf
G(t''-t''')\delta\mathbf\Omega\times\mathbf G(t''')\mathbf
S_-(0)\big]\nonumber\\
&+\big[\mathbf G(t''-t''')\delta\mathbf\Omega\times\mathbf
G(t''')\mathbf S_+(0)\big]\times\big[\mathbf G(t'')\mathbf
S_+(0)\big]\Big\}. \label{dstwo}
\end{align}
The processes corresponding to this equation are that at time $t'''$
the diffence in precession frequencies gives rise to a difference in
orientations of the two pseudospins, then the interatomic
interaction acts at time $t''$ and finally the difference of
precession frequencies leads to a change in the total pseudospin of
the pair of states. After simplification Eq.\ (\ref{dstwo}) becomes
\begin{align}
&\delta \mathbf S_+(t)\nonumber\\
=&-8g_{12}S_1(0)S_2(0)\int_0^t dt' \int_0^{t'} dt''
\int_0^{t''}dt'''\mathbf
G(t-t')\delta\mathbf\Omega\nonumber\\
&\times\mathbf G(t'-t'')
\Big\{\mathbf G(t'')\mathbf e_z\times\big[\mathbf
G(t''-t''')\delta\mathbf\Omega\times\mathbf G(t''') \mathbf
e_z\big]\Big\}. \label{dstwos}
\end{align}

The quantity measured in experiment is the population of atoms in
the ground state, ${\rm P}_g=N/2-\sum_i S_{iz}$, where $N$ is the
total number of atoms. On summing Eq.\ (\ref{dstwos}) over all pairs
of motional states one finds
\begin{align}
\delta {\rm P}_g(t)=C_1\Xi_1+C_3\Xi_3+C_3\Xi_3,
\label{twosz}
\end{align}
where
\begin{align}
\Xi_1=&\sum_{i,j}\frac{4g_{ij}S_i(0)S_j(0)}{\Omega_{xy}^3}(\delta\mathbf\Omega_{ij\parallel})^2\\
\Xi_2=&\sum_{i,j}\frac{4g_{ij}S_i(0)S_j(0)}{\Omega_{xy}^3}(\delta\mathbf\Omega_{ij\perp})^2\\
\Xi_3=&\sum_{i,j}\frac{4g_{ij}S_i(0)S_j(0)}{\Omega_{xy}^3}\mathbf e_z\cdot(\delta\mathbf\Omega_{ij\parallel}\times\delta\mathbf\Omega_{ij\perp})\end{align}
and
\begin{align}
C_1=&\frac{\delta}{6\left(1+\delta^2\right)^4}
\left\{3\left(2+\delta^2\right)\sin^2\left(\tau\sqrt{1+\delta^2}\right)\right.\nonumber\\
&-3\left[\tau^2+\left(3\tau^2-4\right)\delta^2
+2\left(2+\tau^2\right)\delta^4\right]\nonumber\\
&\times\cos\left(\tau\sqrt{1+\delta^2}\right)
+3\left[4\delta^4-4\delta^2-\tau^2\left(1+\delta^2\right)\right]\nonumber\\
&+
\tau\sqrt{1+\delta^2}\left[\tau^2+\left(\tau^2+9\right)\delta^2-6\delta^4\right]\nonumber\\
&\left.\times\sin\left(\tau\sqrt{1+\delta^2}\right)\right\},\\
C_2=&\frac{\delta}{2\left(1+\delta^2\right)^{7/2}}\left\{4\delta^2\sqrt{1+\delta^2}
-4\delta^2\sqrt{1+\delta^2}\right.\nonumber\\
&\left.\times\cos\left(\tau\sqrt{1+\delta^2}\right)
-\sqrt{1+\delta^2}\sin^2\left(\tau\sqrt{1+\delta^2}\right)\right.\nonumber\\
&\left.-\tau\left(-1+\delta^2+2\delta^4\right)\sin\left(\tau\sqrt{1+\delta^2}\right)\right\},\\
C_3=&\frac{2\sin^2\left(\tau\sqrt{1+\delta^2}/2\right)}{\left(1+\delta^2\right)^{5/2}}
\nonumber\\
&\times\left\{-\tau\sqrt{1+\delta^2}
+\sin\left(\tau\sqrt{1+\delta^2}\right)\right\},
\end{align}
with $\delta=\Delta/\Omega_{xy}$, $\tau=\Omega_{xy} t$,
$\mathbf\Omega_{xy}=\overline{\mathbf\Omega}_i-\Delta {\mathbf
e}_z$,
$\delta\mathbf\Omega_{ij\parallel}=(\delta\mathbf\Omega_{ij}\cdot\mathbf\Omega_{xy})
\mathbf\Omega_{xy}/\Omega_{xy}^2$, and
$\delta\mathbf\Omega_{ij\perp}=\delta\mathbf\Omega_{ij}-\delta\mathbf\Omega_{ij\parallel}$.
Here the bar denotes an average over motional states.

\begin{figure}
\includegraphics[width=7cm]{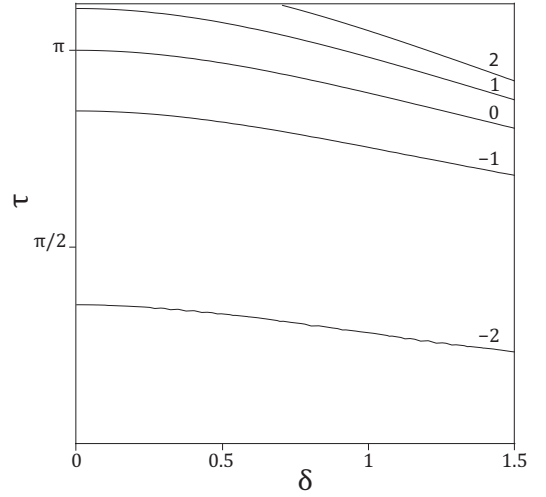}
\vspace{1cm} \caption{Contour plot of the frequency shift
$\delta\omega$, in units of $(2N/\Omega_{xy}^2)\,\overline{ g_{ij}(
\mathbf\Omega_i - \mathbf\Omega_j)^2}$ as a function of
dimensionless detuning, $\delta=\Delta/\Omega_{xy}$ and
dimensionless pulse duration, $\tau =\Omega_{xy} t_p$.} \label{c1}
\end{figure}

Experimentally, the pulse duration $t_p$ is fixed and pairs of
detunings $\Delta_1$ and $\Delta_2$ which render the same population
${\rm P}_g$ in the ground state at the end of the pulse, i.e.,
$S_z(t_p,\Delta_1)=S_z(t_p,\Delta_2)$, are measured.  The clock
shift is defined as $\delta\omega=(\Delta_1+\Delta_2)/2$. The
reason why this method can detect the effects of interatomic interactions lies in
the symmetry of the Bloch equation (\ref{bloch}). For simplicity,
let us assume $\Omega_{iy}=0$. Without interactions,
Eq.~(\ref{bloch}) is invariant under the transformation
$\{\Delta,S_{ix},S_{iy},S_{iz}\}\to\{-\Delta,-S_{ix},S_{iy},S_{iz}\}$,
which implies $\Delta_1=-\Delta_2$ and $\delta\omega=0$. When
$g\neq0$, Eq.~(\ref{bloch}) becomes invariant under the
transformation
$\{\Delta,g,S_{ix},S_{iy},S_{iz}\}\to\{-\Delta,-g,-S_{ix},S_{iy},S_{iz}\}$,
which indicates $\delta\omega\neq0$ if $\Omega_i$ are different. To
order $g(\delta\Omega)^2$, from Eq.~(\ref{twosz}), we have
\begin{align}
\delta\omega=\frac{2\delta
S_z(t_p,\Delta)}{dS_z^0(t_p,\Delta)/d\Delta}.\label{do}
\end{align}

{\it Field inhomogeneity.} To compare our results with experiment we
need to take into account the form of the probe laser field, which
is a linearly polarized Gaussian beam. With the assumption that the
axis of symmetry of the laser field coincides with that of the
optical trap, the electric field vector is given by
\begin{align}
\mathbf E_-(\rr)=&E_0\mathbf
e_x\frac{w(0)}{w(z)}\exp\left(-\frac{x^2+y^2}{w^2(z)}\right)\nonumber\\
&\times\exp\left(i\frac{2\pi}{\lambda_L}
z+i\frac{\pi(x^2+y^2)}{\lambda_L R(z)}-i\zeta(z)\right),
\end{align}
with the field width $w(z)=w(0)\sqrt{1+(z/z_R)^2}$, the radius of
curvature $R(z)=z\left[1+(z_R/z)^2\right]$ and the Gouy phase
$\zeta(z)=\arctan(z/z_R)$. The Rayleigh range is $z_R=\pi
w(0)^2/\lambda_L$. The laser wavelength is $\lambda_L\approx698$ nm
and the beam divergence is estimated to be $\theta=\lambda_L/\pi
w(0)\approx0.01$ \cite{campbell}. Atoms interact with each other
only within a single pancake.  For motional states specified by the quantum numbers for the three Cartesian directions $\mathbf
n_i=(n_{ix},n_{iy},0)$      associated with a given pancake,  the spread
of the directions of $\mathbf\Omega_i$ over the harmonic motional
states  is smaller than $10^{-3}$ in the temperature regime $T\sim
1$ $\mu$K. Therefore we can neglect terms involving
$\delta\mathbf\Omega_{\perp}$ in Eq.~(\ref{twosz}). In
Fig.~(\ref{c1}) we show contours of constant frequency shift as a
function of pulse duration and laser detuning for conditions of
experimental interest.   In agreement with experiment, we find that
the shift can vanish, even though $g$ is nonzero.  However, contrary
to initial expectations  \cite{campbell}, vanishing of the shift
does not correspond to there being equal numbers of atoms in the
ground and excited states at the end of the pulse, ${\rm
P}_g(t_p)=N/2$. From Fig.~(\ref{c1}), for $\tau<\pi$, we see that
for longer pulses, the zero shift occurs at smaller detunings, i.e.,
larger ${\rm P}_g$. As the temperature decreases,  atoms will
concentrate more around the center of the trap, leading to a larger
$\Omega_{xy}$, e.g., longer pulses $\Omega_{xy} t_p$ (for fixed
$t_p$). Thus we predict that as the temperature is lowered,
 the zero shift point will move to larger ${\rm
P} _g$, a result consistent with experiment \cite{campbell}.

The temperature dependence of the magnitude of the frequency shift
can be easily extracted in the regime $\omega_{\perp}\ll T\ll 1/m
w^2(0)$. The coarse-grained density distributions associated with
the two-dimensional harmonic oscillator wave functions have the
semiclassical form $\sim \left[(n_x-m\omega x^2/2)(n_y-m\omega
x^2/2)\right]^{-1/2}$. Thus $g_{ij}\sim
[\min(n_{ix},n_{jx})\min(n_{iy},n_{jy})]^{-1/2}$. The difference in
the driving frequencies
$(\mathbf\Omega_{i}-\mathbf\Omega_{j})^2\sim(n_{ix}+n_{iy}-n_{jx}-n_{jy})^2$.
Since typical values of the $n_x$ and $n_y$ are proportional to $T$, the average of the prefactor $\Xi_1$ in Eq.~(\ref{twosz})  $
\overline{g_{ij}(\mathbf\Omega_{i}-\mathbf\Omega_{j})^2}$ behaves as $T$.
The ratio between the magnitude of the shift measured at $3$ $\mu$K
and that at $1$ $\mu$K is about three \cite{campbell}, in agreement
with this behavior.

In this paper we have derived a closed expression for the dependence of the clock shift on the duration of the probe pulse, the laser detuning, and the geometry of the probe laser beam.   In the calculations we have taken into account the fact that  the effective exchange interaction between atoms in different motional states depends on the specific states in question, and that the ``Rabi frequency" $\mathbf \Omega_i$ depends on the motional state.
As a check on our understanding of the basic physics it would be valuable to confirm the predicted dependences. In particular, one could make experiments for the case when the probe beam is not collinear with the axis of the trap.

We are very grateful for many valuable discussions with Jan Thomsen
on the experiments of Ref.~\cite{campbell}. Useful conversations with Jun Ye are also acknowledged.


\begin{thebibliography}{99}
\bibitem{Directions}  To avoid making the notation too clumsy we denote directions in coordinate space and in pseudospin space by $x$, $y$, and $z$, but the reader should bear in mind that the two spaces are distinct.
\end{thebibliography}
\end{document}